\documentclass[letterpaper, 10pt, conference]{ieeeconf}  

\IEEEoverridecommandlockouts                              
\usepackage{amsmath} 
\usepackage{amssymb}  

\usepackage[pdftex]{graphicx}

\usepackage{url}

\def\Ref#1{(\ref{#1})}

\newcommand{\sgn}{{\rm sgn}}

\newcommand{\cN}{{\mathcal N}}

\newcommand{\nn}{\nonumber}
\newcommand{\al}[1]{\begin{eqnarray}#1\end{eqnarray}}
\newcommand{\eq}[1]{\begin{equation}#1\end{equation}}

\newcommand{\alequal}{&\!\!\!=\!\!\!&}

\newcommand{\alnothing}{&\!\!\! \!\!\!&}

\newcommand{\discFil}{Discrete Path Assignment Filter }
\newcommand{\contFil}{Continuous Path Assignment Filter }

\title{\LARGE \bf
Path Assignment Techniques For Vehicle Tracking
}

\author{Richard Altendorfer
\thanks{R. Altendorfer is with Driver Assistance Systems, 
ZF TRW, Koblenz, Germany
        {\tt\small richard.altendorfer@zf.com}}   
 and Sebastian Wirkert
\thanks{S. Wirkert is with Deutsches Krebsforschungszentrum, Heidelberg,
        Germany
        {\tt\small s.wirkert@dkfz-heidelberg.de}}
}

\begin{document}

\maketitle
\thispagestyle{empty}
\pagestyle{empty}

\begin{abstract}

Many driver assistance systems such as Adaptive Cruise Control require the identification of the closest vehicle that is in the host vehicle's path.
This entails an assignment of detected vehicles to the host vehicle path or neighboring paths.
After reviewing approaches to the estimation of the host vehicle path and lane assignment techniques we introduce two methods that are motivated
by the rationale to filter measured data as late in the processing stages as possible in order to avoid delays and other artifacts of intermediate
filters. These filters generate discrete posterior probability distributions from which a path or ``lane'' index is extracted by a median estimator.
The relative performance of those methods is illustrated by a ROC using experimental data and labeled ground truth data.

\end{abstract}

\section{INTRODUCTION}

Driver assistance systems that react to other vehicles e. g. adaptive cruise control (ACC) or automatic emergency braking (AEB) or 
collision mitigation braking (CMB) need to have a means to identify the vehicle/object that is most relevant to the system under consideration. For ACC for example the object is sought that the host vehicle is following whereas for AEB or CMB the object is sought that
the host vehicle is most likely to crash into. In ACC systems this function is often referred to as track selection or target selection or target object selection (TOS). 

The common approach to tackle this problem has been to obtain a fused and filtered estimate of the host vehicle path based on all available proprioceptive and exteroceptive sensor data. There is ample literature on obtaining an improved estimate of the host vehicle path by fusing and filtering inertial data (wheel speeds, yaw rate) with video line markings, road boundaries, common motion of other vehicles, and other clues about the host vehicle's path (HVP), see e.~g. \cite{Eidehall_Gustafsson_2004,Lane_data_fusion_Sparbert_2004,Polychronopoulos_et_al_2007}.
After having obtained an HVP estimate the closest object inside this path is selected as ACC target. This approach works well if the estimated path is not very noisy and in particular does not veer over several lane widths on a time scale much shorter than that of a normal lane change.
However, if the path estimate exhibits fast fluctuations which can happen e. g. if the path is only based on inertial data, the path assignment might fail occasionally. Modifying the path filter such that those fast fluctuations will be suppressed may also make the track selection less sensitive in situations where a fast object selection or deselection is required such as in lane changes of the host vehicle or the target vehicle. 
Instead we propose methods that postpone the filtering which is commonly applied to the host vehicle path to the final quantity, namely the lane assignment itself.

The topic of path or lane assignment has already been addressed in the following articles:
In \cite{Lane_Tracking_Rong_Li_2008} the vehicle's lane states are modeled by a three-state Hidden Markov Model to which the Viterbi algorithm to estimate the optimal lane assignment {\it sequence} as well as a so-called {\it Lane Filter} to recursively estimate the optimal {\it current} lane assignment is applied. This {\it Lane Filter} turns out to be a discrete Bayes filter followed by a MAP estimator. 

In \cite{Target_selection_Koreans_10} the lane assignment is performed by fuzzy logic in the lateral position and lateral velocity in circular path coordinates from which an unfiltered quantity called {\it lane-probability} is derived.

In \cite{Schubert_Wanielik_2010} the normalized lateral position (NLP) is used as a continuous generalization of the discrete lane number. 
This NLP is computed by propagation of probabilities using the unscented transform and is based upon the target vehicle's position in Cartesian host vehicle coordinates and (possibly fused) lane curvature and lane width information. The discrete lane assignment is performed using postulated lane occupancy likelihood functions and contains no filtering.

In \cite{Schiffmann_Schwartz_patent_2011} two methods -- a discrete and a continuous one -- of a probabilistic lane assignment are described. 
The continuous method describes the computation of a continuous lane ``number" similar to the NLP from  \cite{Schubert_Wanielik_2010}. This continuous lane number is computed at every time step by propagation of probability distributions to second order and is not filtered.
The discrete method computes the probabilities of the host vehicle being in three different lanes as well as the probabilities of the target vehicle being in those lanes and then computes their joint probabilities by multiplication thereby implicitly assuming statistical independence.
Again, those joint probabilities are not filtered.
Both methods use an approximation of a circular arc by a third order polynomial and also approximate the lateral position of the target objects perpendicular to the estimated curved path by the lateral position with respect to a host vehicle fixed Cartesian coordinate system.

In \cite{Mueller_Pauli_et_al_2011} an incremental sampling of the probability distributions of target vehicles and the host vehicle path was applied to the non-linear equation for the {\it time to cutin/cutout}.

In all of the above described approaches (except \cite{Lane_Tracking_Rong_Li_2008}) filtering is not performed on the lane index or the NLP but occurs at
earlier stages of the estimation. In contrast we propose here two filters where the filtering is applied to the final quantity - the path index - or the lateral position in path coordinates which is related to the path index by algebraic discretization. 
In the first technique presented in this paper we apply a discrete Bayes filter to the lane index - 
similar to the {\it Lane Filter} from \cite{Lane_Tracking_Rong_Li_2008} - albeit with adaptive measurement likelihoods that take into account the uncertainties of all measured quantities and a velocity-dependent Markov transition matrix with a band structure.

This paper is organized as follows: in the first section we illustrate the concept of a host vehicle path and introduce the
two new methods to be compared as well as the comparison setup. Then we describe the two path assignment approaches, namely the \discFil and the \contFil and their corresponding estimators. In the fourth section we
present experimental results of the comparison of the two approaches based on an unfiltered path estimate using yaw rate and host vehicle speed. A review of the probability distribution transformation from Cartesian to circular path coordinates is relegated to the appendix.

\section{Comparison setup}  

\begin{figure}[ht]
\centering
\includegraphics[viewport = 1cm 0.5cm 20cm 18cm, clip, width = 0.8 \columnwidth]{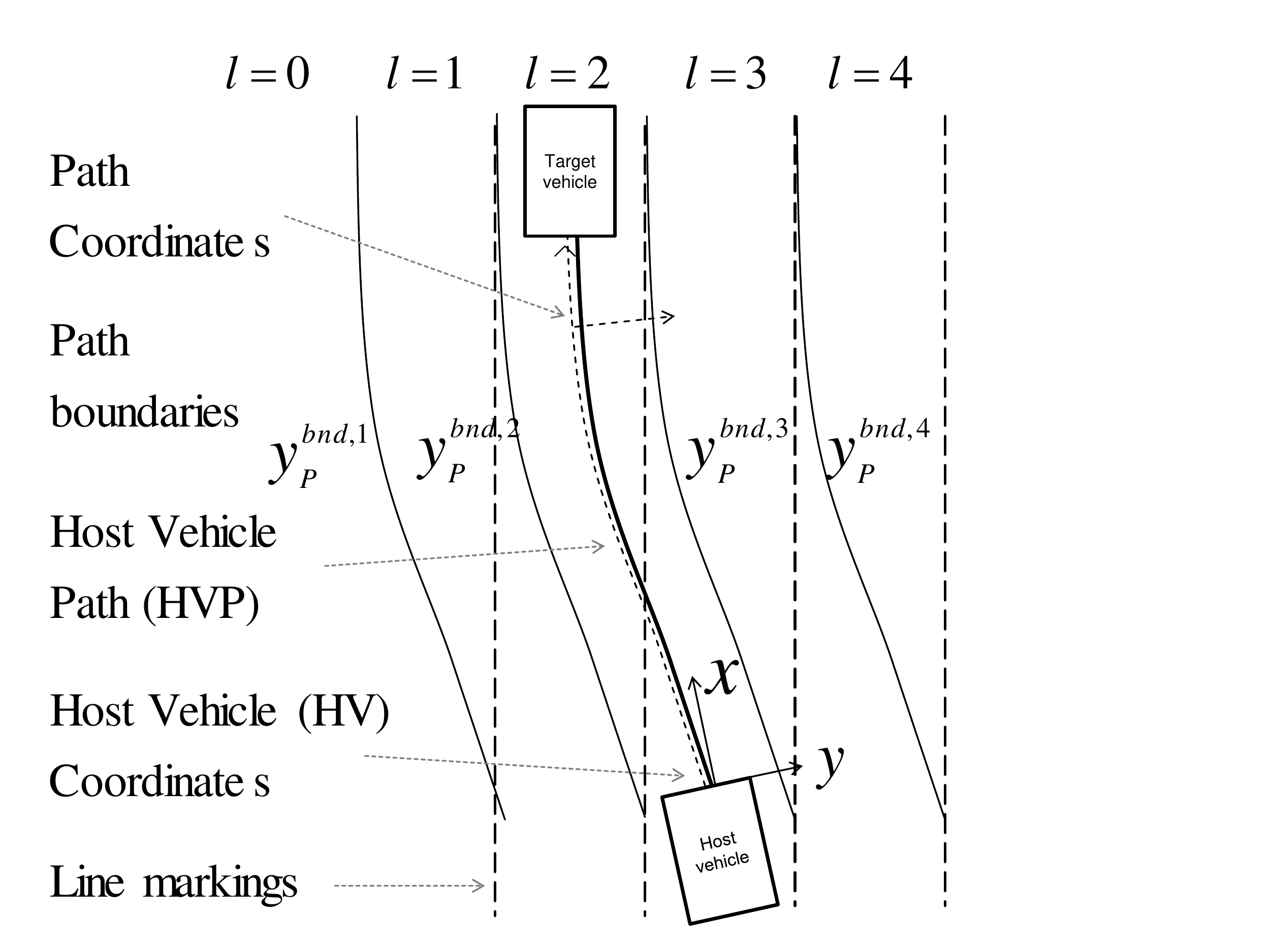}
\caption{Bird's eye view of host and target vehicle with host vehicle coordinates and host vehicle path coordinates.}
\label{fig_trajectory_coordinates}
\end{figure}

We want to assign vehicles/objects ahead of the host vehicle either to its path or its neighboring ``lanes''. Note that the host vehicle path need not coincide with traffic lanes delimited by line markings; this is the case e. g. in lane change scenarios, see fig. \ref{fig_trajectory_coordinates}. Here we also distinguish between the host vehicle (HV) coordinates $(x,y)$ which describe a Cartesian, vehicle fixed coordinate system and the host vehicle path (HVP) coordinates $(x_P,y_P)$ which describe a locally orthogonal coordinate system 
whose origin is also at the host vehicle and whose curved $x-$axis follows the HVP.

The variables $\{y^{bnd,i}_{P}\}$ with $i=1\dots 4$ delimit the host vehicle path and neighboring ``lanes'' along the HVP.
We want to estimate the HVP index $l\in L = \{0, 1, 2, 3, 4\}$ for all detected vehicles; its interpretation is given in table \ref{laneInterpretation}.
\begin{table}[ht]
\label{laneInterpretation}
\caption{Path index}
\centering
\begin{tabular}{c||c}
\hline
$l$ & interpretation \\
\hline\hline
$0$ & {left of left path} \\
\hline
$1$ & {left path} \\
\hline
$2$ & {center (host) path} \\
\hline
$3$ & {right path} \\
\hline
$4$ & {right of right path}\\
\hline
\end{tabular}
\end{table}

\begin{figure}[ht]
\centering
\includegraphics[viewport = 0.1cm 1.5cm 25.5cm 17.5cm, clip, width = 1 \columnwidth]{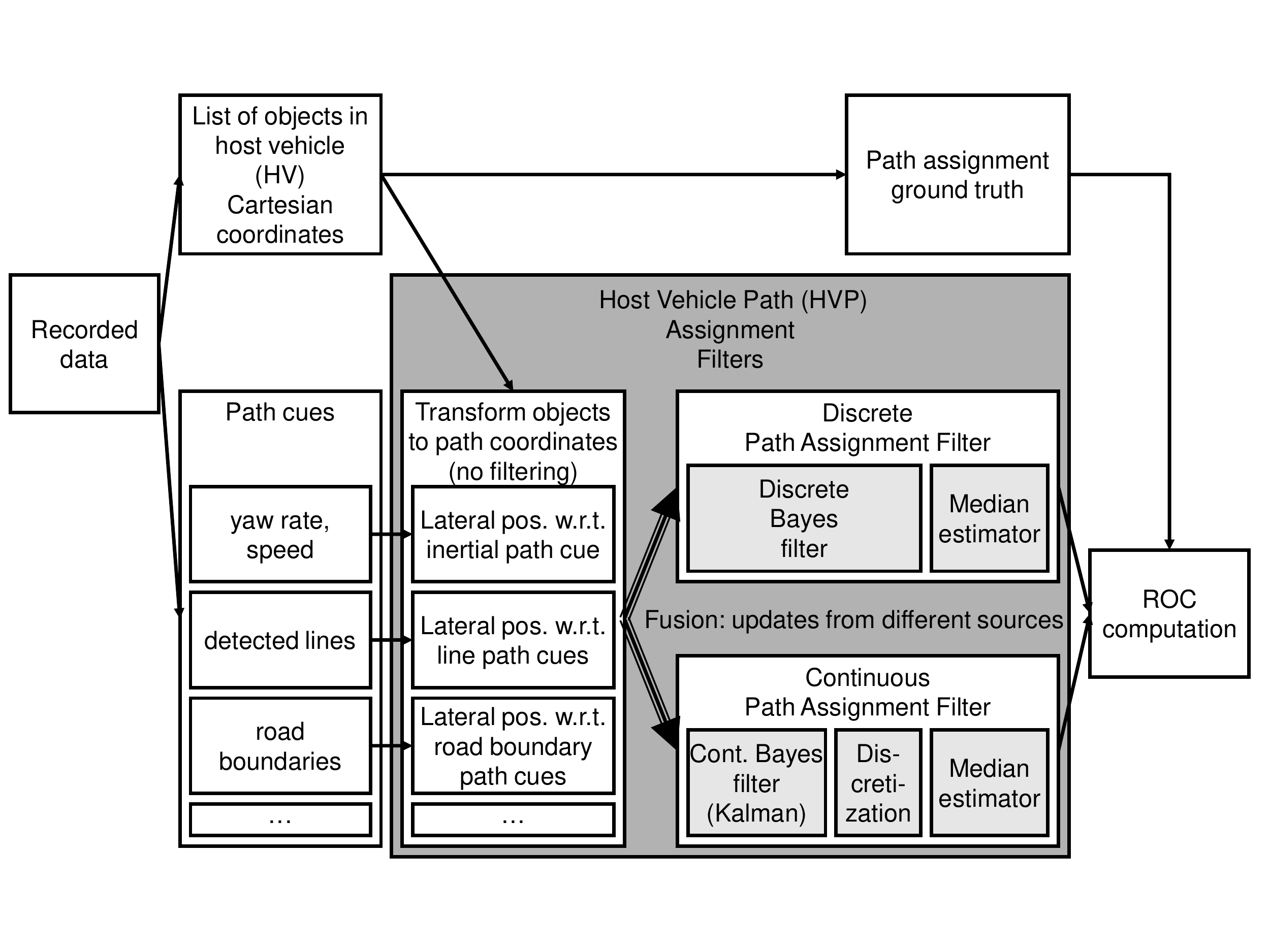}
\caption{Overview of the HVP assignment filter and the simulation setup to compare the discrete and continuous approaches to path assignment.}
\label{fig_setup_comparison}
\end{figure}

As an input to the HVP assignment filter we assume a list of vehicle objects - possibly based upon multi-sensor fusion - with kinematic attributes $(x,y,\dot x,\dot y)$ in Cartesian vehicle fixed coordinates. We also assume sensor input with cues about the vehicle path: e. g. an ``inertial" path based upon yaw rate and speed (see also the appendix) as well as line markings detected by video. These individual path cues are then fed into the HVP assignment filter module depicted in dark gray in fig. \ref{fig_setup_comparison}. The first stage in the HVP assignment filter module is the transformation of the objects' probability distribution functions (pdfs) from HV coordinates to HVP coordinates using the individual path cues, also represented by pdfs.
The crucial point here is that this transformation as well as the original path cue sensor inputs should not contain any filters in order to postpone filtering to the final stage. In particular, yaw rate, speed, detected lines, etc, should be raw measurements.
The objects' pdfs in HVP coordinates are then fed into two different filters: one is a discrete Bayes filter with a discrete posterior path index distribution as output, the other is a continuous Bayes filter represented by a Kalman filter with a posterior lateral path coordinate distribution as output. This continuous distribution is discretized using the measurement likelihood function
of the discrete Bayes filter. Then median estimators are applied to both distributions in order to arrive at an estimate of the path index.

In the next two sections the two filter choices - the discrete and the continuous Bayes filter - are introduced.

\section{\discFil}

The probability of an assigned lane $l$ at time $t_k$ depends on the history of the lateral positions of objects, i.~e. $p(l_k|Z_k)$ where $Z_k = \{z_{P,k}, z_{P,k-1}, \dots\}$. Here, $z_{P,k}$ denotes the set of measurements at time $t_k$ containing the lateral position of the object in path coordinates $y^{obj}_{P,k}$ and the set of measured lane boundaries in path coordinates $\{y^{bnd,1}_{P,k}, y^{bnd,2}_{P,k}, y^{bnd,3}_{P,k}, y^{bnd,4}_{P,k}\}$: 
\eq{
z_{P,k} =\{y^{obj}_{P,k},y^{bnd,1}_{P,k}, y^{bnd,2}_{P,k}, y^{bnd,3}_{P,k}, y^{bnd,4}_{P,k}\}.
\label{eq_composite_measurement}
}
In many situations only the two innermost boundaries $y^{bnd,2}_{P,k}, y^{bnd,3}_{P,k}$ will be detected by a video system and the position of the other two boundaries will then have to be extrapolated from the width of the inner lane. If no lane boundary is detected then also default values for the innermost boundaries will have to be used.

To this filtering problem a discrete Bayes filter can be applied, for a review see e. g. \cite{Thrun05,Altendorfer_Matzka_12}.

\subsection{Bayes filter update}

The update step of a discrete Bayes filter for this setup reads
\eq{
p(l_k|Z_k) = { p(z_{P,k}|l_k)p(l_{k}|Z_{k-1}) \over \sum_{m_k \in L} p(z_{P,k}|m_k) p(m_k|Z_{k-1}) } \label{eq_Bayes_update_lane_likelihoods}
}
Hence we need to model the measurement likelihood $p(z_{P,k}|l_k)$.

\subsubsection{Modeling of measurement likelihoods}

The Bayes update requires an expression for the measurement likelihood $p(z_{P,k}|l_k)$. However, it is not straightforward to model this likelihood since the state $l_k$ with five discrete values is much simpler than the set of measurements $z_{P,k}$. Therefore we will specify the inverse measurement model $p(l_k|z_{P,k})$ and then relate it to the measurement likelihood by using Bayes' theorem.

By Bayes' theorem the measurement likelihood is related to the inverse measurement model by 
\eq{
p(z_{P,k}|l_k) = p(l_k|z_{P,k}) {p(z_{P,k}) \over p(l_k) } \label{eq_Bayes_theorem_lane_likelihood}
}
Assuming a uniform a priori distribution for $p(l_k)$ the quotient in eq. \Ref{eq_Bayes_theorem_lane_likelihood} cancels in the Bayes update expression and eq. \Ref{eq_Bayes_update_lane_likelihoods} becomes
\eq{
p(l_k|Z_k) = { p(l_k|z_{P,k})p(l_{k}|Z_{k-1}) \over \sum_{m_k \in L} p(m_k|z_{P,k}) p(m_k|Z_{k-1}) } 
}

To compute the probability that lane $l$ is occupied at $t_k$ we compute the probability that the object is left of the right boundary $y^{bnd,l+1}_{P,k}$ and subtract the probability that the object is left of the left lane boundary $y^{bnd,l}_{P,k}$, i.~e.
\al{
p(l_k|z_{P,k}) &\!\!\!\!\!=\!\!\!\!\!& \int_{-\infty}^{0} p(y^{obj}_{P,k} -y^{bnd,l+1}_{P,k}| z_{P,k} )d(y^{obj}_{P,k} - y^{bnd,l+1}_{P,k}) \nn\\ 
               &\!\!\!\!\! \!\!\!\!\!\!\!\!& - \int_{-\infty}^{0} p(y^{obj}_{P,k}  -y^{bnd,l}_{P,k}| z_{P,k} )d(y^{obj}_{P,k} - y^{bnd,l}_{P,k})  \label{eq_lane_occupancy_prob}
}
If the individual probability distributions for the lateral positions in HVP coordinates of the object and the lane boundaries are represented by $\cN(y^{obj}_{P,k}; \mu^{obj}_{P,k}, \sigma^{obj}_{P,k})$ and $\cN(y^{bnd,l}_{P,k}; \mu^{bnd,l}_{P,k}, \sigma^{bnd,l}_{P,k})$ the integrands in eq. \Ref{eq_lane_occupancy_prob} above are given by
\al{
&&\hspace{-9mm}p(\tilde y^{bnd,l}_{P,k}| z_{P,k} )=\nn\\
&&\cN(\tilde y^{bnd,l}_{P,k}; \mu^{obj}_{P,k} - \mu^{bnd,l}_{P,k}, \sqrt{(\sigma^{obj}_{P,k})^2 + (\sigma^{bnd,l}_{P,k})^2})\nn
}
where $\tilde y^{bnd,l}_{P,k} := y^{obj}_{P,k} - y^{bnd,l}_{P,k}$. Then
\al{
p(l_k|z_{P,k}) \alequal \Phi\left( { \mu^{bnd,l+1}_{P,k} - \mu^{obj}_{P,k} \over \sqrt{(\sigma^{obj}_{P,k})^2 + (\sigma^{bnd,l+1}_{P,k})^2}} \right)\nn\\
\alnothing - \Phi\left( { \mu^{bnd,l}_{P,k} - \mu^{obj}_{P,k} \over \sqrt{(\sigma^{obj}_{P,k})^2 + (\sigma^{bnd,l}_{P,k})^2}} \right)\label{eq_lane_occ_prob_final}
}
where $\Phi$ is the cumulative distribution function of the standard normal distribution.
By using the inverse measurement model we have derived the Bayes update step under the mild assumption of a uniformly distributed a priori lane assignment probability. 
This approach differs from the one in \cite{Schubert_Wanielik_2010} where the measurement model $p(z_{P,k}|l_k)$ was postulated in a piecewise, heuristic manner.

\begin{figure}[ht]
\centering
\includegraphics[viewport = 3cm 9cm 18cm 20.5cm, clip, width = 0.87 \columnwidth]{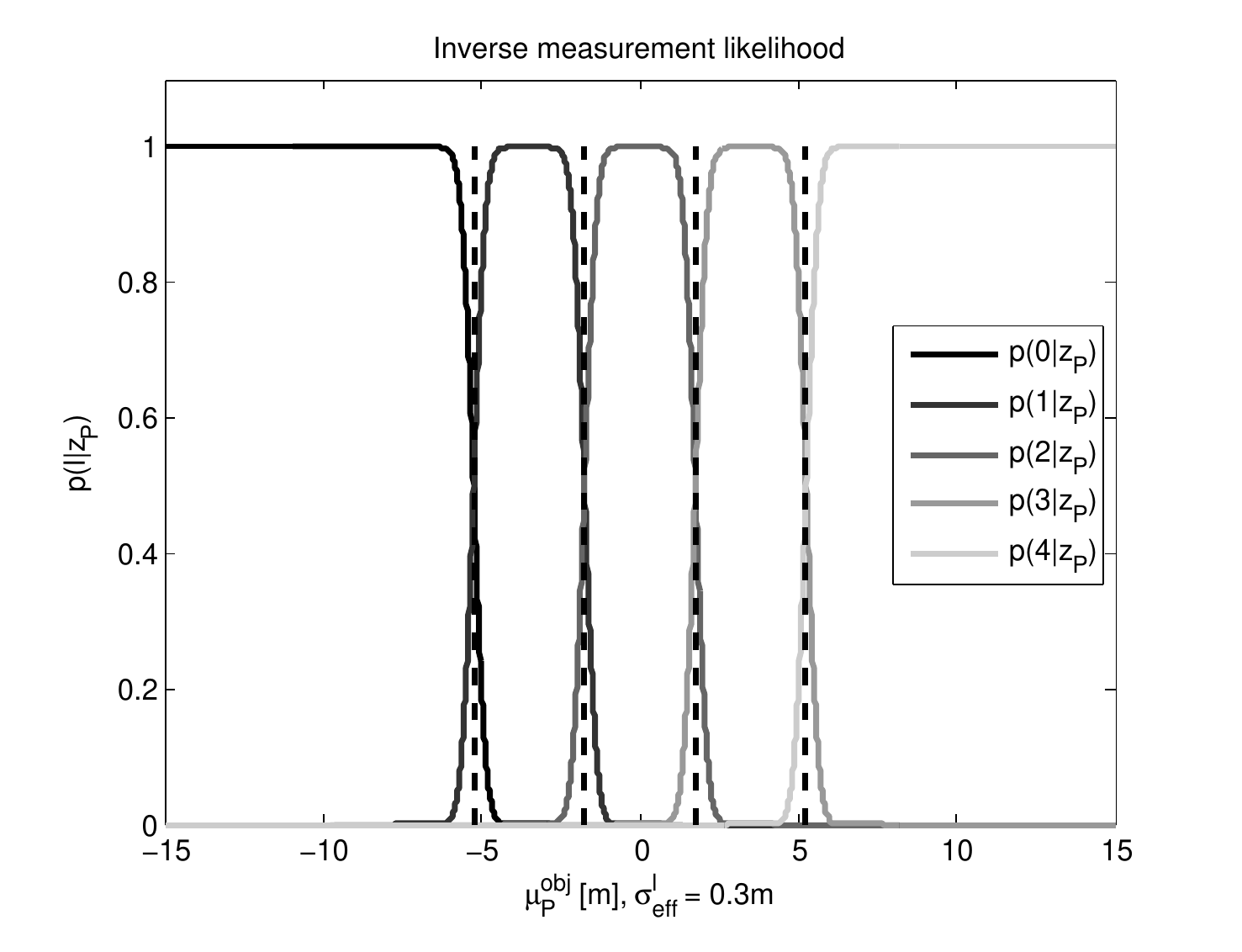}
\caption{Inverse measurement likelihood functions for $\sigma^l_{eff} = 0.3 m$.}
\label{fig_inverse_measurement_likelihoods}
\end{figure}

\begin{figure}[ht]
\centering
\includegraphics[viewport = 3cm 9cm 18cm 20.5cm, clip, width = 0.87 \columnwidth]{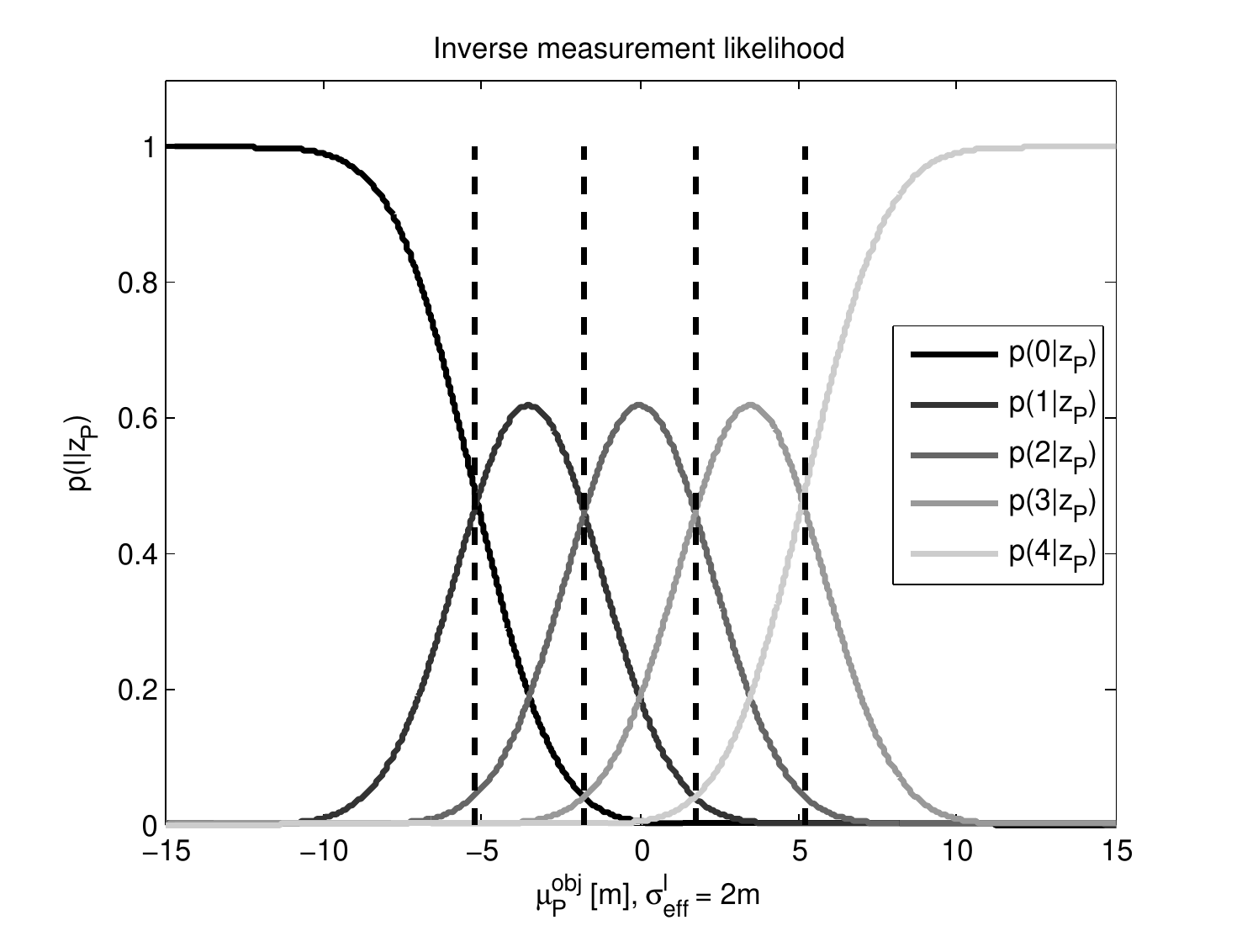}
\caption{Inverse measurement likelihood functions for $\sigma^l_{eff} = 2.0 m$.}
\label{fig_inverse_measurement_likelihoods_2m}
\end{figure}

The parameters $\mu^{bnd, l}_{P,k}, \sigma^{bnd,l}_{P,k}$ constitute input from the transformation module (``Transform objects to path coordinates'') in fig. \ref{fig_setup_comparison}.
As an illustration the inverse measurement likelihoods have been plotted in fig. \ref{fig_inverse_measurement_likelihoods} and fig. \ref{fig_inverse_measurement_likelihoods_2m} for two different values of the effective variance $(\sigma^l_{eff,k})^2=(\sigma^{obj}_{P,k})^2 + (\sigma^{bnd,l}_{P,k})^2$ where the variances of all lane boundaries were taken to be identical.

It can be seen that for a small effective variance the lane assignment likelihoods stay roughly constant over almost the entire width of the lane whereas for large effective variances the likelihoods approximate a Gaussian shape except for the boundary likelihoods. Integration of the likelihood functions over $\mu^{obj}_{P,k}$ yields the width of the lane which is infinite for the likelihoods of the boundary states 'left of left path' and 'right of right path'.

\subsection{Bayes filter prediction}

The Bayes filter prediction is given by 
\eq{
p(l_{k}|Z_{k-1}) = \sum_{l_{k-1}} p(l_{k}|l_{k-1}) p(l_{k-1}|Z_{k-1})
}
where the Markov transition matrix $p(l_{k}|l_{k-1})$ is modeled as a perturbation of the identity matrix, i.~e. objects tend to stay in their lanes, with small probabilities for transitions to neighboring lanes, but not to next-to-neighboring lanes.\footnote{A similar transition structure albeit with partly different states was used in \cite{Paffrath_10}.}
\al{
&&\hspace{-8mm}p(l_{k}|l_{k-1}) = \label{eq_Markov_matrix}\\
&&\hspace{-8mm}\everymath{\scriptstyle}\begin{pmatrix}
1 - \epsilon - \eta	& \epsilon + 0.5|\eta| - \eta	& 0 												& 0 												& 0 									\cr 
\epsilon + \eta 			& 1 - 2\epsilon - |\eta| 		& \epsilon + 0.5|\eta| - \eta	& 0 												& 0 									\cr 
0 									& \epsilon + 0.5|\eta| + \eta & 1 - 2\epsilon - |\eta| 		& \epsilon + 0.5|\eta| - \eta	& 0 									\cr 
0 									& 0 												& \epsilon + 0.5|\eta| + \eta	& 1 - 2\epsilon - |\eta| 		& \epsilon - \eta 			\cr 
0 									& 0													& 0 												& \epsilon + 0.5|\eta| + \eta	& 1 - \epsilon + \eta  
\end{pmatrix} \nn
}
Here, $\epsilon$ denotes the default probability for transitions between neighboring lanes and $\eta$ is an additional increment or decrement parameter, respectively, that can depend for example upon the lateral velocity of the object or upon a detected indicator signal. The specific expressions involving $\eta$ arise from the following assumptions: the velocity perturbation is additive; a velocity to the right should increase the transition
probability from the current lane the right lane and decrease the probability to the left path; and the
columns of the Markov transition matrix must add up to one for proper normalization.

\subsection{Estimator for HVP assignment}\label{sec_estimator_discrete_case}

\begin{figure}[ht]
\centering
\includegraphics[viewport = 3cm 9.5cm 18cm 20.5cm, clip, width = 0.66 \columnwidth]{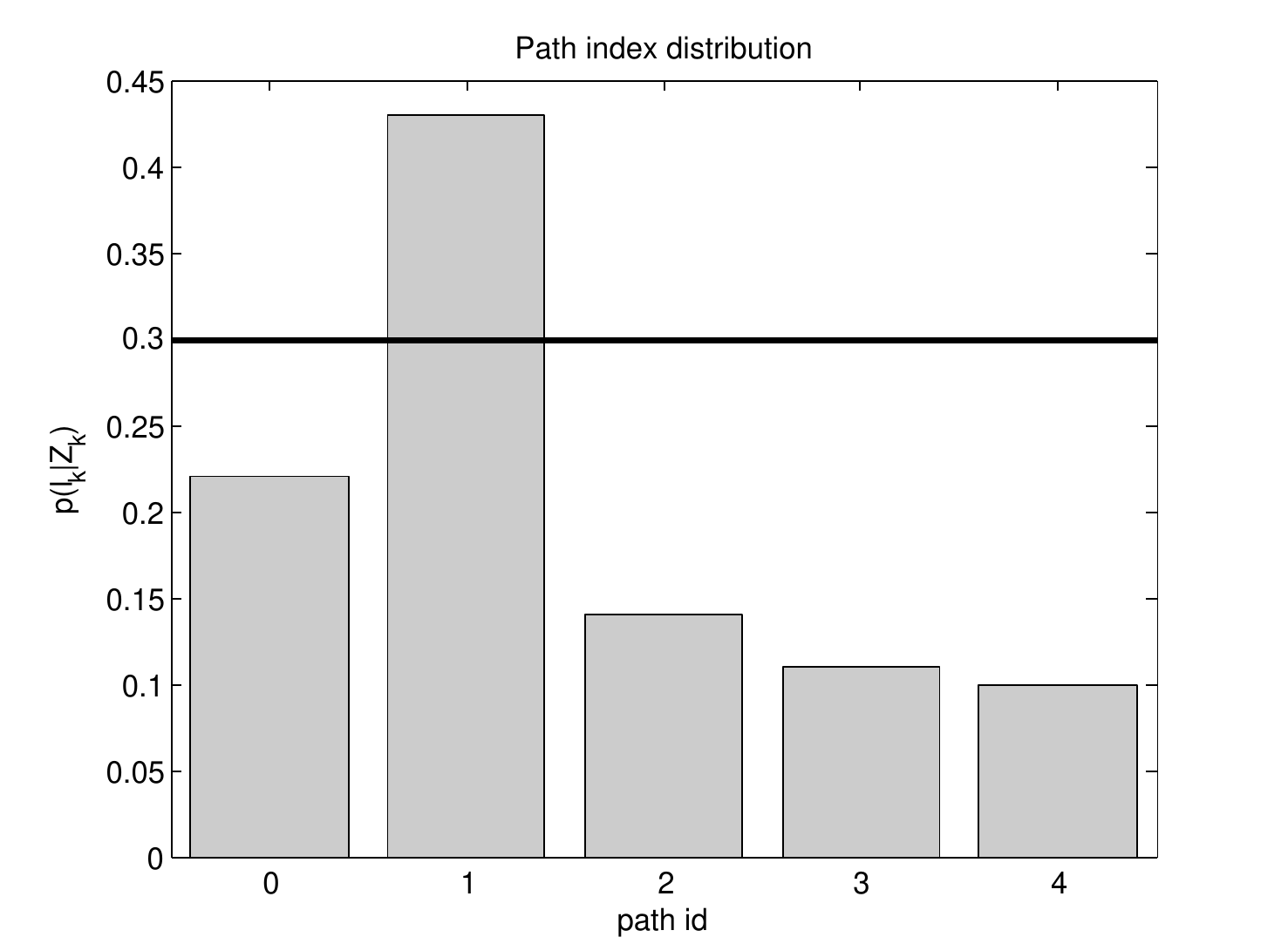}
\caption{Path probability distribution where the left path (path id = 1) is assigned since it is the median of the distribution and exceeds a threshold of 0.3.}
\label{fig_Lane_probability_distribution}
\end{figure}

After having obtained an a posteriori probability distribution of lane indices by discrete Bayes filtering a single lane index estimate 
needs to be extracted by an estimator. This is done by a median estimator, i. e. by determining the lane index $l_k^{est}$ for which
\eq{
\sum_{l_k \leq l_k^{est}} p( l_k | Z_k ) \geq 0.5 \ \textnormal{ and } \sum_{l_k \geq l_k^{est}} p( l_k | Z_k ) \geq 0.5
}
If the probability of this lane estimate is also above a certain threshold $p( l_k^{est} | Z^k ) \geq p^l_{min}$ then this estimate is accepted as an assigned lane for use by a subsequent track selection; otherwise no lane will be assigned to this object. We have decided against a mean estimator because it is less robust and against a MAP estimator because for strongly oscillating host vehicle paths the lane probability distribution can exhibit several maxima of similar height. An example distribution where the left lane is assigned is depicted in fig. \ref{fig_Lane_probability_distribution}.

\section{\contFil}

The first step in the continuous case is the filtering of the object's lateral position in HVP coordinates: $y^{obj}_P$.
Then, its continuous distribution function is mapped onto a discrete lane assignment distribution using path boundary information.
Finally, a median estimator is applied to extract a single lane assignment value.

\subsection{Dynamical system for lateral HVP coordinate}

The state is $\hat y^{obj}_P$ where the diacritic\ $\hat{ }$\ serves to distinguish this filtered quantity from 
the unfiltered measurement $y^{obj}_P$ in eq. \Ref{eq_composite_measurement}. In order to be able to compare results using continuous filtering to 
discrete filtering on an equal footing the velocity is modeled as an input $u_k$ to the dynamical system in analogy to 
\Ref{eq_Markov_matrix} and is not part of the state. However, inclusion of derivatives (velocity, acceleration, ...) of the lateral position
in the state is straightforward and appropriate for more dynamic situations.
The dynamical model is a discrete-time counterpart white noise velocity model with
\eq{
\xi_{k+1} = \xi_{k} + \Delta t_k u_k  + \Delta t_k \nu_{k} \label{eq_dynamicalModel}
}
with $\Delta t_k = t_{k+1} - t_k$ and $\nu_{k}$ the process noise.
The measurement equation is
\eq{
z_{k} = \xi_{k} + w_{k}
}
where $z_{k}$ is the lateral position in host vehicle path coordinates. For circular paths arising from host vehicle yaw rate and velocity, the lateral position is given by eq. \Ref{eq_y_HVP_coordinate} as described in the appendix. For non-circular paths, this lateral position can be provided by other path cues as indicated in fig. \ref{fig_setup_comparison}. To this one-dimensional system a Kalman filter is applied yielding a posterior distribution $\cN(\hat y^{obj}_{P,k}; \hat\mu^{obj}_{P,k}, \hat\sigma^{obj}_{P,k})$. In order to satisfy one of the prerequisites for a Kalman filter we have also verified in the appendix that the propagation of pdfs of the parameters of eq. \Ref{eq_y_HVP_coordinate} by Taylor expansion is a valid approximation; hence $z_{k}$ is given by eq. \Ref{eq_composite_measurement_mean} from the appendix and the variance of the measurement noise $w_{k}$ is given by eq. \Ref{eq_composite_measurement_var}.

\subsection{Map to discrete lane distribution and estimator}

With the lane boundary information contained in $z_{P,k}$ (from eq. \Ref{eq_composite_measurement}) the inverse measurement likelihood eq. \Ref{eq_lane_occ_prob_final} can be applied to arrive at a discrete probability distribution for the five path assignment possibilities from table \ref{laneInterpretation}. Note that
in this case the mean and standard deviation $\mu^{obj}_{P,k}, \sigma^{obj}_{P,k}$ are replaced by $\hat\mu^{obj}_{P,k}, \hat\sigma^{obj}_{P,k}$ from the filtered pdf of $\hat y^{obj}_{P,k}$.
Finally, as in sec. \ref{sec_estimator_discrete_case} a median estimator is applied followed by thresholding.

\section{Experimental results}

\begin{figure}[ht]
\centering
\includegraphics[viewport = 3.5cm 9.5cm 18cm 20.5cm, clip, width = 1.1 \columnwidth]{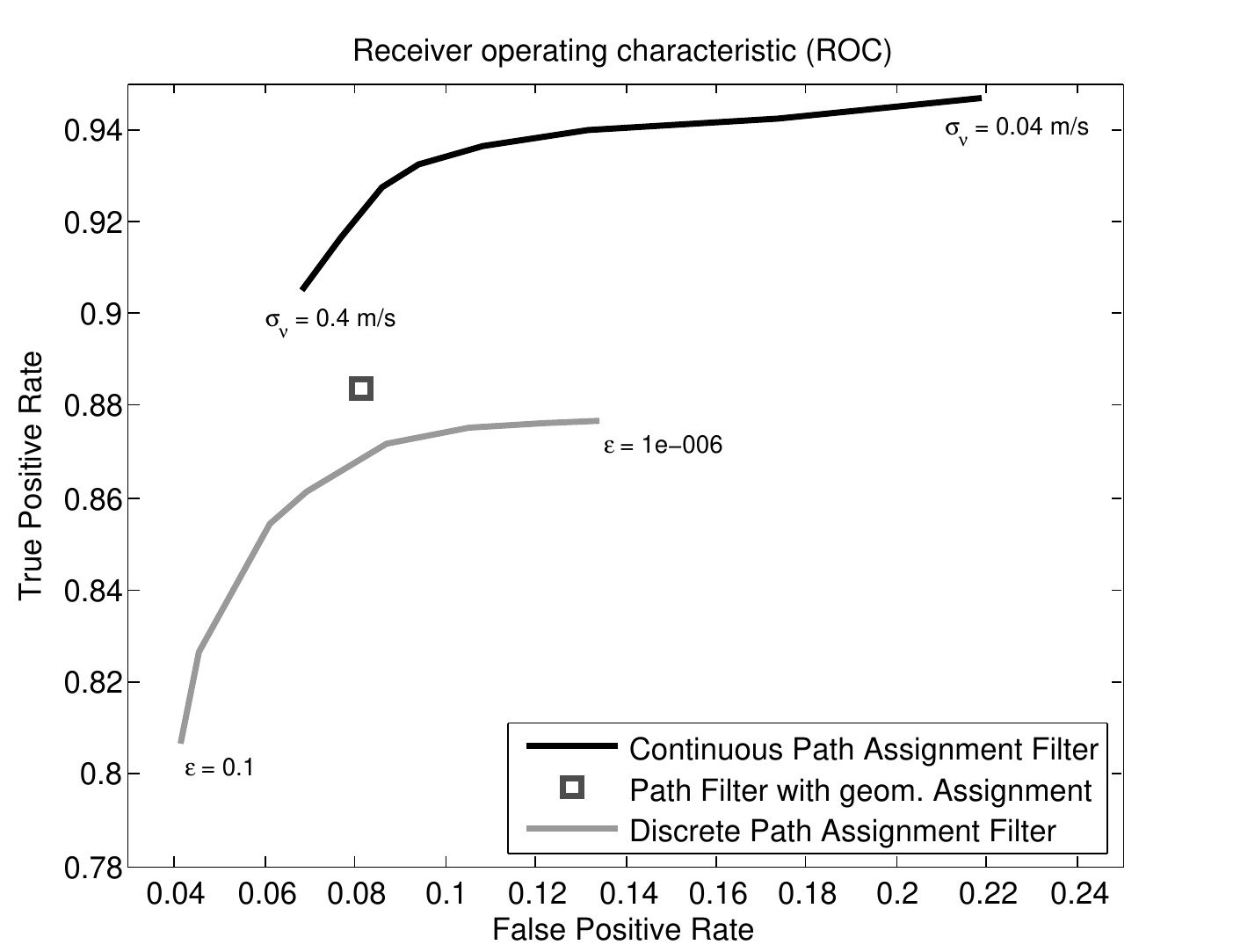}
\caption{ROC curve of correct host lane assignments (true positives) versus incorrect host lane assignments (false positives).}
\label{fig_ROC_lane_assignment}
\end{figure}

We have evaluated both algorithms using recorded experimental data from highway and country road scenarios with a total length of ca. 117 minutes of ground truthed data. Here path cue inputs from the camera were disabled, hence the only source of path information was due to the inertial host vehicle signals yaw rate and speed. For a representative comparison of both approaches we have reduced the number of free parameters in each algorithm to one: namely the value of the default transition to neighboring lanes $\epsilon$ for the \discFil and the standard deviation $\sigma_\nu$ of the process noise for the Continuous Path Assignment Filter. 
Other parameters such as sensor characteristics or the thresholds for the median estimators are identical. With those parameters in the ranges $\epsilon \in \{1e^{-1}...1e^{-6}\}$ and $\sigma_\nu \in \{0.04...0.4m/s\}$ the ROC curve in fig. \ref{fig_ROC_lane_assignment} was generated.

It can be seen that the \contFil has a better true positive (TP) rate than the \discFil whereas the false positive (FP) rate can be lower for the \discFil albeit at the cost of a low TP rate. For comparison with a path assignment method where filtering is done at the path estimation level and not at the lane assignment level we have also plotted the ROC point of an assignment method where the path itself is filtered and estimated and the path assignment is done by projecting the vehicles' position into this path and the neighboring paths and evaluating the assigned path geometrically using its current lateral position (``Path Filter with geom. Assignment").

The overall worse performance of the discrete filter might be due the fact that the discrete filter
contains less information than the continuous one because the discrete probability distribution consists of five values for each lane or path. Hence information whether the target vehicle is directly in the middle of a lane or displaced to the left or right boundary is averaged over by a single probability for this lane. In contrast, the continuous filter generates a continuous a posteriori distribution for the lateral displacement with respect to the host vehicle trajectory $y^{obj}_{P,k}$. The subsequent discretization is just an algebraic map without additional filtering.

\section{Conclusions}

We have comparatively evaluated two techniques of path assignment that were motivated by the requirement to postpone filtering as late as possible in the estimation process. In the discrete case the filtering was indeed performed on the final discrete path index distribution, in the continuous case the filtering was performed on the lateral displacement followed by discretization. We have derived path index measurement likelihoods - used by both techniques - from first principles making the minimal assumption of a uniform a priori distribution.

Our numerical study suggests that the continuous filtering although the filtering is not directly done on the final quantity has advantages with respect to the discrete approach. Furthermore, in addition to the numerical evidence in favor of the continuous filtering, the discrete approach contains more modeling assumptions: the assignment probability is averaged over the width of a lane in the filtering stage - as discussed above, and the prediction requires modeling assumptions about lane transitions which were subsumed in this paper into the parameters $\epsilon$ and $\eta$.
We have also shown that in relevant parameter regions of forward looking sensors the propagation of probability distributions by Taylor expansion of formula \Ref{eq_y_HVP_coordinate} is a valid approximation.






\section{Appendix: Inertial HVP transformation}\label{app_inertial_HVP_transformation}

\begin{figure}[ht]
\centering
\includegraphics[viewport = 6cm 2.5cm 18cm 15cm, clip, width = 0.55 \columnwidth]{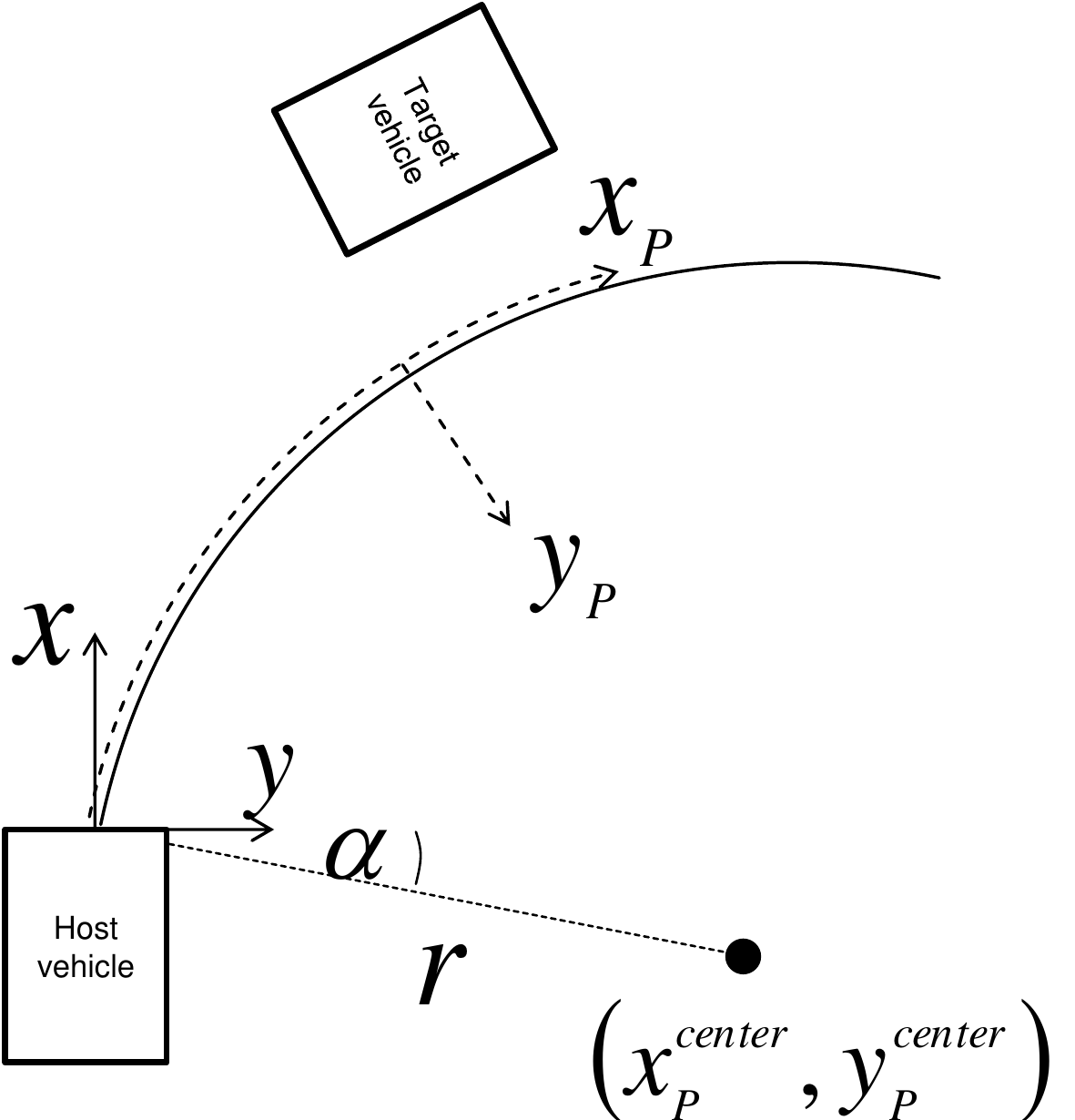}
\caption{Bird's eye view of host and target vehicle with host vehicle coordinates and host vehicle trajectory coordinates from inertial data.}
\label{fig_inertial_trajectory_coordinates}
\end{figure}

If no lane markings are detected and no other information about the HVP such as guard rails are available, then the only source
of information about the HVP are inertial host vehicle data, i. e. the host vehicle speed $v$ and its yaw rate $\dot\psi$. Using the common assumption of a steady-state circular motion the vehicle will travel on 
a circular arc with radius $r = {v \over \dot\psi}$ as depicted in fig. \ref{fig_inertial_trajectory_coordinates}.

\begin{figure}[ht]
\centering
\includegraphics[viewport = 0.5cm 6cm 21cm 22cm, clip, width = 0.8 \columnwidth]{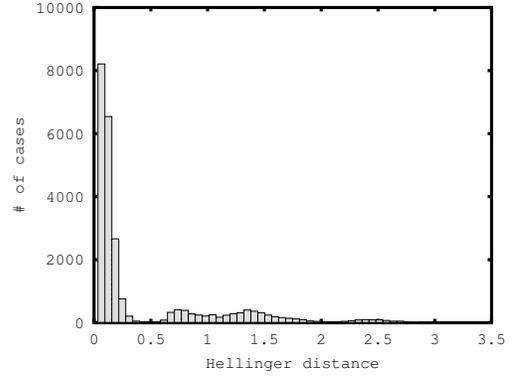}
\caption{Distribution of Hellinger distances.}
\label{fig_distri_hellinger}
\end{figure}

Here we also allow for a non-vanishing heading angle $\alpha$, i.~e. the host vehicle path at the host vehicle's reference point -- the middle of the front bumper -- is not tangential to the vehicle's longitudinal axis but is rotated by an angle $\alpha$
\eq{
\begin{pmatrix} x^{center}_{p} \cr y^{center}_{p} \end{pmatrix} = \begin{pmatrix} \cos(\alpha) & -\sin(\alpha) \cr \sin(\alpha) & \cos(\alpha) \end{pmatrix} \begin{pmatrix} 0 \cr r \end{pmatrix} \label{eq_center_coordinates}
}
Then the lateral distance $y_P$ in HVP coordinates $(x_P,y_P)$
of an object with HV coordinates $(x,y)$ is given by (see also \cite{Target_selection_Koreans_10,Schubert_Wanielik_2010})
\eq{
y_P = r - \sgn(r)\sqrt{\left(x + \sin(\alpha) r\right)^2 + \left(y - \cos(\alpha) r\right)^2}   \label{eq_y_HVP_coordinate}
}
This is the minimal distance to the estimated HVP.
In a probabilistic framework the parameters $\mu_{P,k}, \sigma_{P,k}$ of the pdf of $y_{P,k}$ must be determined. The pdfs of the signals in eq. \Ref{eq_y_HVP_coordinate} are propagated by Taylor expansion, i.~e. given the vector $\zeta_k = ({v}_{k}, {\dot\psi}_{k}, {x}_{k}, {y}_{k})^\top$, its mean $\mu_{\zeta_k} = (\mu^{v}_{k}, \mu^{\dot\psi}_{k}, \mu^{x}_{k}, \mu^{y}_{k})^\top$ and its covariance matrix $V_{\zeta_k}$ we get
\al{
\mu_{P,k} \alequal {\mu^{v}_{k} \over \mu^{\dot\psi}_{k}} - \sgn({\mu^{v}_{k} \over \mu^{\dot\psi}_{k}})\cdot \label{eq_composite_measurement_mean}\\
        \alnothing \cdot \sqrt{(\mu^{x}_{k}+ \sin(\alpha){\mu^{v}_{k} \over \mu^{\dot\psi}_{k}})^2 + (\mu^{y}_{k} - \cos(\alpha){\mu^{v}_{k} \over \mu^{\dot\psi}_{k}})^2} \nn \\
\sigma^2_{P,k} \alequal  \partial_{\zeta} y_{P,k}\left( \mu_{\zeta_k} \right) V_{\zeta_k} \left( \partial_{\zeta} y_{P,k}\left( \mu_{\zeta_k} \right) \right)^\top \label{eq_composite_measurement_var}
}
Here, $\partial_{\zeta} y_{P,k}\left( \mu_{\zeta_k} \right)$ denotes the Jacobi matrix of $y_{P,k}$ at $\mu_{\zeta_k}$.
In order to validate that Taylor expansion results in an accurate approximation of the pdf of eq. \Ref{eq_y_HVP_coordinate} we 
have performed a Monte-Carlo study where 24576 different pdfs of $x, y, v,$ and $\dot\psi$ were propagated through \Ref{eq_y_HVP_coordinate} by sampling with 5000 samples. The values for $x$ ranged between one and $110m$, the value for $y$ was calculated by letting the angle range between $-21$ and $21$ degrees. Velocities of $1$ to $70 m/s$ were taken into account and the yaw rates considered ranged from $-0.7$ to $0.7 rad/s$. 
Then we compared the resulting normalized histograms to Gaussian distributions parametrized by eqs. \Ref{eq_composite_measurement_mean} and \Ref{eq_composite_measurement_var}. For comparison of two pdfs the Hellinger distance was chosen since it satisfies all conditions of a mathematical metric.
In fig. \ref{fig_distri_hellinger} the distribution of Hellinger distances between the Monte-Carlo run and the Gaussian from Taylor expansion is plotted for all 24576 pdfs.
 
\begin{figure}[ht]
\centering
\includegraphics[viewport = 0.5cm 6cm 21cm 22cm, clip, width = 0.8 \columnwidth]{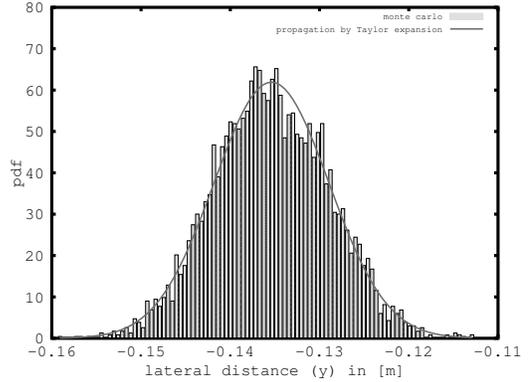}
\caption{Monte Carlo result versus Gaussian from Taylor expansion with the largest Hellinger distance of 3.02.}
\label{fig_max_hellinger}
\end{figure}

The largest distance value is ca. $3$. In order to assess the consequence of such a value we have plotted the normalized histogram with the largest Hellinger distance together with its corresponding Gaussian in fig. \ref{fig_max_hellinger}. It can be seen that the histogram with that distance still approximates the Gaussian to high accuracy. Hence we conclude that in the chosen parameter ranges the propagation of pdfs by Taylor expansion is appropriate.
 
\bibliographystyle{IEEEtran}
\bibliography{bibliography}

\end{document}